\UseRawInputEncoding
\documentclass{cup-hpl}

\usepackage{lipsum}
\usepackage{graphicx}
\usepackage{overpic}

\begin{document}

\shorttitle{SRS in the degenerate regime}
\shortauthor{Yao Zhao et al.}

\title{Stimulated Raman scattering in the degenerate regime}

\author[1]  {Yao Zhao \corresp{Yao Zhao. \email{yaozhao@siom.ac.cn}}}
\author[2,3]{Suming Weng}
\author[2,3,4]{Zhengming Sheng}
\author[1,3]{Jianqiang Zhu}

\address[1]{Key Laboratory of High Power Laser and Physics, Shanghai Institute of Optics and Fine Mechanics, Chinese Academy of Sciences, Shanghai 201800, China}
\address[2]{Key Laboratory for Laser Plasmas (MoE), School of Physics and Astronomy, Shanghai Jiao Tong University, Shanghai 200240, China}
\address[3]{Collaborative Innovation Center of IFSA (CICIFSA), Shanghai Jiao Tong University, Shanghai 200240, China}
\address[4]{SUPA, Department of Physics, University of Strathclyde, Glasgow G4 0NG, UK}

\begin{abstract}
Stimulated Raman scattering (SRS) in plasma in the degenerate regime is studied theoretically and numerically. Different from normal SRS with the non-degenerate eigen electrostatic mode excited, the degenerate SRS is developed at plasma density $n_e>0.25n_c$ when the laser amplitude is larger than a certain threshold. To satisfy the phase-matching conditions of frequency and wavenumber, the excited electrostatic mode has a constant frequency around half of the incident light frequency $\omega_0/2$, which is no longer the non-degenerate eigenmode of electron plasma wave $\omega_{pe}$. Both the scattered light and the electrostatic wave are trapped in plasma with their group velocities being zero. Super hot electrons are produced by the degenerate electrostatic wave. Our theoretical model is validated by particle-in-cell simulations. The SRS driven in this degenerate regime is an important laser energy loss mechanism in the laser plasma interactions as long as the laser intensity is higher than $10^{15}$W/cm$^2$.
\end{abstract}


\maketitle

\section{Introduction}

Laser plasma interactions (LPI) are widely associated with many applications, such as inertial confinement fusion (ICF) \cite{Campbell2017Laser,froula2010experimental,myatt2014multiple}, radiation sources \cite{Liu2019}, plasma optics \cite{lancia2016signatures,lehmann2013nonlinear}, and laboratory astrophysics \cite{Drake,falk2018}. The concomitant parametric instabilities found in LPI are nonlinear processes which can greatly affect the outcome \cite{Gibbon}. Generally, laser plasma instabilities \cite{kruer1988physics,montgomery2016two}, especially stimulated Raman scattering (SRS), stimulated Brillouin scattering (SBS) and two-plasmon decay (TPD) instability, have been mainly considered in ICF with the incident laser intensity less than $10^{15}$W/cm$^2$ \cite{craxton2015direct,Lindl2014Review,moody2001backscatter}. However, the laser intensity may be in the order of $10^{16}$ or even $10^{17}$W/cm$^2$ in shock ignition \cite{Betti2016Inertial,Batani2014,cristoforetti2019,Klimo2010Particle,gu2019}, Brillouin amplification \cite{Weber2005,lancia2010experimental}, and the interactions of high power laser with matter \cite{rethfeld2017modelling,price1995absorption,george2019}. Therefore, the parametric instabilities close to the regime of subrelativistic intensity needs to be explored in depth.

As well known, SRS usually develops in plasma density not larger than the quarter critical density $n_e\leq 0.25n_c$ due to the decay of the scattering light in its propagation in the overdense density plasma \cite{Liu2019,kruer1988physics}. In the density region $n_e\leq0.25n_c$, the electrostatic wave is the non-degenerate eigenmode of the electron plasma wave. Relativistic intensity lasers can reduce the effective electron plasma frequency, and therefore non-degenerate eigenmode SRS may develop at $n_e>0.25n_c$ \cite{zhao2014effects}. In this work, we show the presence of degenerate SRS, which is found at plasma density $n_e>0.25n_c$ even without considering the relativistic effect. The development of degenerate electrostatic mode is described by the linear perturbations of fluid equations, which may lead to a few subsequent nonlinear phenomena \cite{Weber2005,Ghizzo2006,wu195202,riconda2006two}. This mode develops only when the laser intensity exceeds a certain threshold. The theoretical model is supported by particle-in-cell (PIC) simulations.

\section{Theoretical analysis of SRS in the degenerate regime}

Generally, SRS is a three-wave instability that a laser decays into an electrostatic wave with frequency equal to the eigen electron plasma wave, and a light wave. However, the stimulated electrostatic wave is no longer the eigenmode of the electron plasma wave in the SRS degenerate regime, where both the frequencies of scattered light and electrostatic field are nearly half of the incident laser frequency. The mechanism of this instability can be described by the SRS dispersion relation at plasma density $n_e>0.25n_c$.

To investigate the degenerate SRS mechanism in laser plasma interactions, we firstly introduce the nonrelativistic dispersion relation of SRS in cold plasma \cite{kruer1988physics}
\begin{equation}
\omega_e^2-\omega_{pe}^2=\frac{\omega_{pe}^2k_e^2c^2a_0^2}{4}\left(\frac{1}{D_{e+}}+\frac{1}{D_{e-}}\right),
\end{equation}
where $D_{e\pm}=\omega_e^2-k_e^2c^2\mp2(k_0k_ec^2-\omega_0\omega_e)$, and $a_0$ is the laser normalized amplitude. The relation between laser intensity $I$ and $a_0$ is given by $I(\mathrm{W}/\mathrm{cm}^2)=1.37\times 10^{18}a_0^2/[\lambda(\mu \mathrm{m})]^2$. Also in Eq. (1), $\omega_0$ and $\omega_e$ are the frequencies of incident laser and electrostatic wave, respectively. $k_0$ and $k_e$ are respectively the wavenumbers of pump laser and electrostatic wave. Generally, we have Re$(\omega_e)=\omega_{pe}$ in the SRS non-degenerate eigenmode regime $n_e\leq0.25n_c$. However, when the amplitude of incident laser $a_0$ larger than a threshold, stimulated degenerate electrostatic mode Re$(\omega_e)\neq\omega_{pe}$ will be developed at $n_e>0.25n_c$.

\begin{figure}
\centering
    \begin{tabular}{lc}
        \begin{overpic}[width=0.5\textwidth]{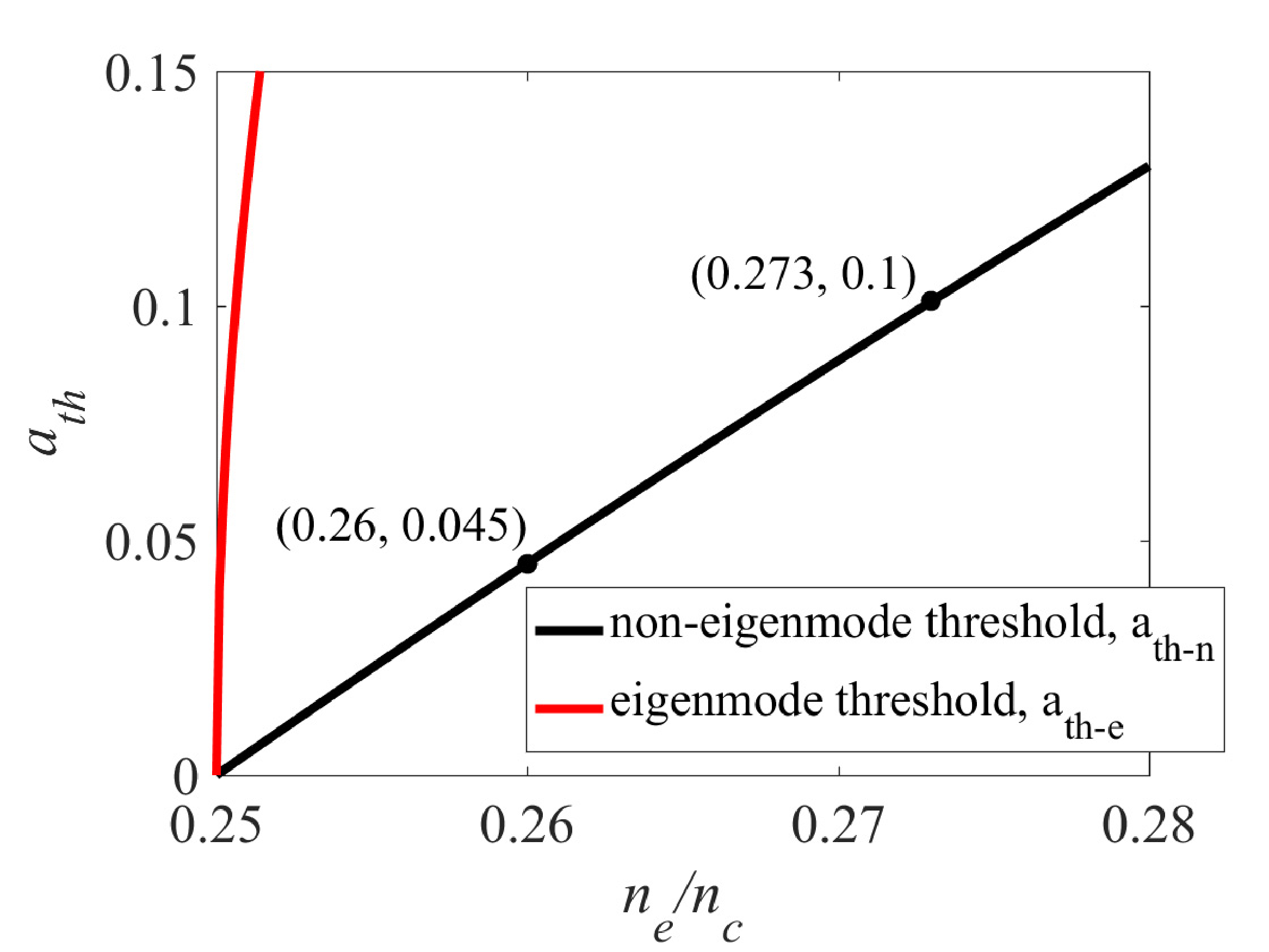}
        \end{overpic}
    \end{tabular}
\caption{ Amplitude thresholds for the development of non-degenerate eigenmode and degenerate SRS in plasma above the quarter critical density. The threshold for the case of non-degenerate eigenmode SRS $a_{th-e}$ is due to the relativistic effect.
    }
\end{figure}

Now we analytically solve Eq. (1) under $n_e>0.25n_c$. Let $\omega_e=\omega_{er}+i\omega_{ei}$, where $\omega_{er}$ and $\omega_{ei}$ are the real and imaginary part of $\omega_e$, respectively. The wavenumber of scattering light is a real $k_sc=0$ in the degenerate regime, i.e., the scattered light is trapped in the plasma. And to keep the phase-matching conditions, we set the electrostatic wavenumber $k_ec=k_0c$. In this case, the imaginary part of Eq. (1) can be simplified to
\begin{equation}
\begin{split}
&(\omega_0\omega_{ei}-2\omega_{ei}\omega_{er})(\omega_{ei}^2-\omega_{er}^2-2\omega_0\omega_{er}+3\omega_0^2\\
&-3\omega_{pe}^2)(\omega_{er}^2+\omega_{ei}^2-\omega_{er}\omega_0-\omega_{pe}^2)=0.\\
\end{split}
\end{equation}
Equation (2) is satisfied for any $\omega_{pe}$ when $\omega_{er}=\omega_0/2$. Therefore, the frequency of the electrostatic wave is a constant, and is independent of the plasma density. The phase velocity of the electrostatic wave is around $v_{ph}=\omega_{er}/k_e\gtrsim c/\sqrt{3}\sim0.58c$.

\begin{figure}
\centering
    \begin{tabular}{lc}
        \begin{overpic}[width=0.5\textwidth]{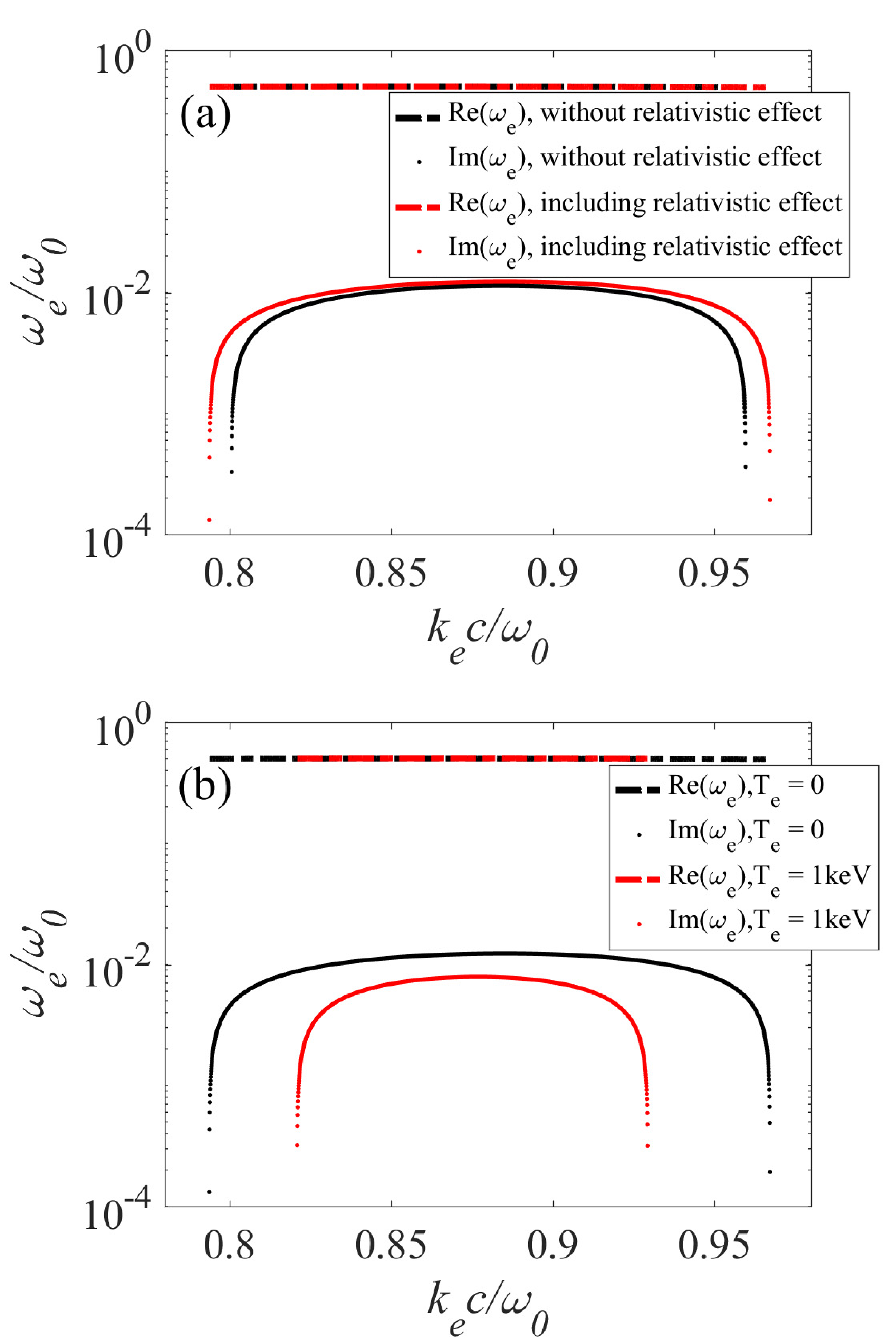}
        \end{overpic}
    \end{tabular}
\caption{ Numerical solutions of SRS dispersion equation at plasma density $n_e=0.27n_c$ with laser amplitude $a_0=0.1$. (a) The relativistic modification on the degenerate SRS at $T_e=0$. (b) The effect of electron temperature on degenerate SRS. The dotted line and dashed line are the imaginary part and the real part of the solutions, respectively.
    }
\end{figure}

Substituting $\omega_{er}=\omega_0/2$ into the real part of Eq. (1), one obtains the growth rate of degenerate SRS
\begin{equation}
\omega_{ei}=\frac{1}{2}\sqrt{4\omega_{pe}(\omega_0-\omega_{pe})+\omega_{pe}^2a_0^2k_0^2c^2/\omega_0^2-\omega_0^2}.
\end{equation}
The above equation indicates that the growth rate $\omega_{ei}$ is reduced by the increasing of plasma density. The threshold $a_{th-n}$ for SRS developing in the degenerate regime can be obtained from $4\omega_{pe}(\omega_0-\omega_{pe})+\omega_{pe}^2a_{th-n}^2k_0^2c^2/\omega_0^2-\omega_0^2\gtrsim0$, i.e.,
\begin{equation}
a_{th-n}\gtrsim\frac{\omega_0\sqrt{\omega_0^2+4(\omega_{pe}^2-\omega_{pe}\omega_0)}}{\omega_{pe}k_0c}.
\end{equation}
Equation (4) indicates that $0.25n_c$ is the turning point between non-degenerate eigenmode SRS and degenerate SRS, where the threshold $a_{th-n}=0$.

For the density region just near the quarter critical density $n_e\gtrsim0.25n_c$, Eq. (4) can be simplified to $a_{th-n}\gtrsim(8/\sqrt{3})(n_e/n_c-0.25)$. Therefore, the condition for the excitation of degenerate SRS in a plasma with density $n_e\gtrsim0.25n_c$ is that the amplitude of pump laser satisfies the above condition, which is almost linearly proportional to the plasma density.

In the following, we consider the relativistic modification of the SRS degenerate in hot plasma. The dispersion of SRS under the relativistic intensity laser is \cite{Gibbon,zhao2014effects}
\begin{equation}
\omega_e^2-\omega_L^2=\frac{\omega_{pe}'^2k_e^2c^2a_0^2}{4\gamma^2}\left(\frac{1}{D_{e+}}+\frac{1}{D_{e-}}\right),
\end{equation}
where $\omega_L^2=\omega_{pe}'^2+3k_e^2v_{th}^2$ with $\omega_{pe}'=\omega_{pe}/\sqrt{\gamma}$, $\gamma=(1+a_0^2/2)^{1/2}$ and $v_{th}$ are the relativistic factor and electron thermal velocity, respectively. Different from degenerate SRS, the threshold for non-degenerate eigenmode SRS developing in cold plasma with $n_e>0.25n_c$ is $\omega_{pe}'\leq0.5\omega_0$, i.e., $a_{th-e}\geq\sqrt{2(16n_e^2/n_c^2-1)}$. As a comparison, the driven amplitudes for degenerate SRS $a_{th-n}$ and non-degenerate eigenmode SRS $a_{th-e}$ at different plasma densities are shown in Fig. 1. One finds that the amplitude threshold for non-degenerate eigenmode SRS is much larger than degenerate SRS, i.e., $a_{th-e}\gg a_{th-n}$. Therefore, the intensity of SRS in $n_e>0.25n_c$ is underestimated according to the previous eigenmode model. As an example, the threshold for laser driving degenerate SRS at plasma density $n_e=0.26n_c$ is around $a_{th-n}=0.045$. A laser with amplitude $a_0=0.1$ can develop degenerate SRS in the plasma region with density $0.25n_c<n_e\lesssim0.273n_c$.

Following the similar steps of the non-relativistic case, the imaginary part of Eq. (5) is simplified to
\begin{equation}
\begin{split}
&(\omega_0\omega_{ei}-2\omega_{ei}\omega_{er})(\omega_{ei}^2-\omega_{er}^2-2\omega_0\omega_{er}+3\omega_0^2\\
&-3\omega_{pe}'^2)\left(\omega_{er}^2+\omega_{ei}^2-\omega_{er}\omega_0-\omega_L^2\right)=0.\\
\end{split}
\end{equation}
We obtain the same identical relation for the real part $\omega_{er}=\omega_0/2$ from Eq. (6). Note that the relativistic factor and electron temperature has no effect on $\omega_{er}$ which is no longer the non-degenerate eigen frequency $\omega_L$. The dispersion relation of the degenerate electrostatic mode satisfies
\begin{equation}
\omega_{er}=\frac{\omega_0}{2}=\frac{1}{2}\sqrt{(k_e^2c^2+\omega_{pe}^2)}.
\end{equation}
From Eq. (7) we know that the group velocity of degenerate electrostatic wave is $v_g=\delta\omega_{er}/\delta k_e\approx0$. Therefore, electrostatic wave will be trapped in the plasma.

\begin{figure*}
\centering
    \begin{tabular}{lc}
        \begin{overpic}[width=0.78\textwidth]{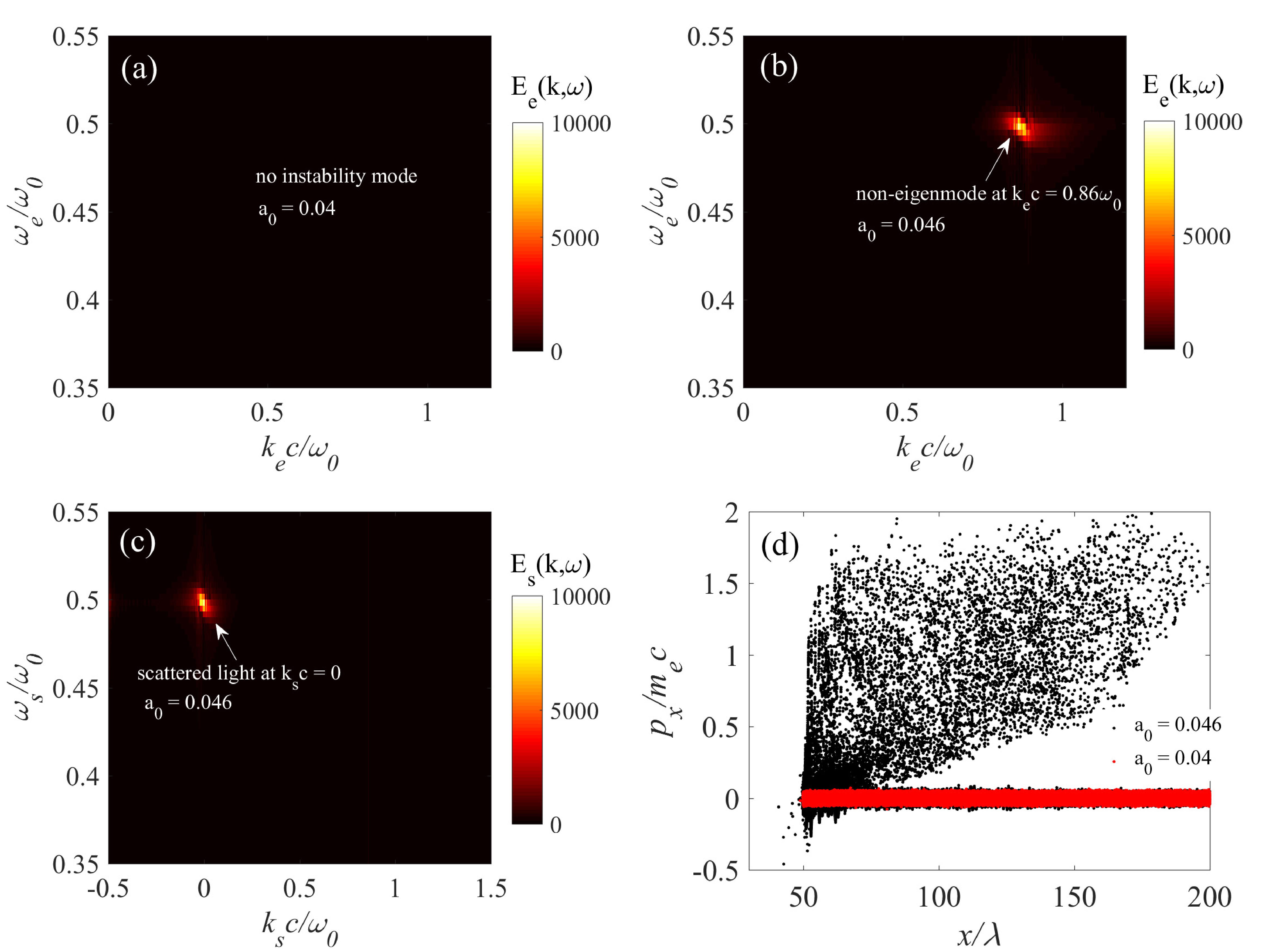}
        \end{overpic}
    \end{tabular}
\caption{ Distributions of the electrostatic wave in $(k_e, \omega_e)$ space obtained for the time window $[100, 400]\tau$ at plasma density $n_e=0.26n_c$ under (a) pump laser amplitude $a_0=0.04$ and (b) pump laser amplitude $a_0=0.046$. (c) Distribution of the electromagnetic wave in $(k_s, \omega_s)$ space obtained under the same conditions of (b). (d) Longitudinal phase space distribution of electrons under different laser amplitudes at $t=600\tau$.
    }
\end{figure*}

The comparisons between the numerical solutions of Eqs. (1) and (5) are exhibited in Fig. 2. One finds that Re$(\omega_e)=\omega_0/2$ is a constant even including relativistic and temperature effects. The frequency of electron plasma wave is reduced by the relativistic factor $\omega_{pe}'=\omega_{pe}/\sqrt{\gamma}$. Therefore, the growth rate $\omega_{ei}$ is increased by the relativistic modification as shown in Fig. 2(a). On the contrary, the frequency of electron plasma wave is enhanced by the electron temperature $\omega_L=\sqrt{\omega_{pe}'^2+3k_e^2v_{th}^2}$, and therefore we find a decrease of the growth rate at higher temperature $T_e=1$keV in Fig. 2(b). Note that the above studies are discussed in the weak relativistic regime, where the plasma density modulation induced by the laser ponderomotive force is weak.

Phase-matching conditions are satisfied in the SRS degenerate regime, therefore the frequency of concomitant light is also Re$(\omega_s)\approx0.5\omega_0$, which can be obtained from the dispersion relation of scattered light
\begin{equation}
\omega_s^2-k_s^2c^2-\omega_{pe}^2=D_{s+}+D_{s-},
\end{equation}
where $D_{s\pm}=\omega_{pe}^2(k_s\pm k_0)^2c^2a_0^2/4[(\omega_s\pm\omega_0)^2-\omega_{pe}^2]$.

According to the linear parametric model of inhomogeneous plasma, the Rosenbluth gain saturation coefficient for convective instability is $G=2\pi\Gamma^2/v_sv_pK'$ \cite{rosenbluth1972}, where $\Gamma$, $v_s$ and $v_p$ are instability growth rate, group velocity of scattering light and plasma wave, respectively. $K$ is the wavenumber mismatch for incident light, scattering light and plasma wave. As it is known, convective instability transits to absolute instability when $K=0$ \cite{liu1974raman}. Based on the above discussions, the mismatching term of degenerate SRS is $K_{\mathrm{ne}}=k_0-k_e-k_s=0$ due to $k_e=k_0$ and $k_s=0$ all the time. Therefore, degenerate SRS is an absolute instability in inhomogeneous plasma.

In conclusion, different from normal SRS, a new type of degenerate SRS can develop in plasma with density $n_e>0.25n_c$. The stimulated electrostatic mode has an almost constant frequency around half of the incident light frequency $\omega_0/2$, which is no longer the non-degenerate eigenmode of the electron plasma wave $\omega_{pe}$. The group velocities of concomitant light and electrostatic wave are zero in the degenerate regime. The degenerate SRS develops only when the laser intensity is higher than a certain threshold which is related to the plasma density.

\section{Simulations for degenerate SRS excitation}
\subsection{1D simulations for degenerate SRS in homogeneous plasma}

To validate the analytical predictions for degenerate SRS, we have performed several one-dimensional (1D) simulations by using the {\sc OSIRIS} code \cite{fonseca2002osiris,hemker2015particle}. The space and time given in the following are normalized by the laser wavelength in vacuum $\lambda$ and the laser period $\tau$. A linearly-polarized semi-infinite pump lasers with a uniform amplitude is incident from the left boundary of the simulation box. In this subsection, only the fluid property of the instability is considered, therefore we set electron temperature $T_e=100$eV with immobile ions. The plasma density is $n_e=0.26n_c$.

Based on Eq. (4) and Fig. 1 we know that the triggering threshold for degenerate SRS is $a_{th-n}=0.045$ at density $n_e=0.26n_c$. To validate the theoretical threshold, two simulation examples under different laser intensities are displayed here. Figure 3(a) shows the case when the laser amplitude is less than the threshold ($a_0=0.04<0.045$), and no instability mode can be found. When the laser amplitude is increased to $a_0=0.046>0.045$, the degenerate electrostatic mode can be found at $k_ec\approx0.86\omega_0$ and $\omega_e\approx0.499\omega_0$ in Fig. 3(b). The corresponding electromagnetic mode with $k_sc\approx0$ and $\omega_s\approx0.5\omega_0$ is shown in Fig. 3(c). These simulation results agree well with the analytical prediction. As discussed above, the phase velocity of the degenerate electrostatic wave is around $v_{ph}\sim0.58c$ at $n_e=0.26n_c$. Therefore, numbers of electrons are heated enormously at the nonlinear stage $t\gtrsim600\tau$ in the SRS degenerate regime as compared to the case below the threshold as shown in Fig. 3(d).

\begin{figure*}
\centering
    \begin{tabular}{lc}
        \begin{overpic}[width=0.98\textwidth]{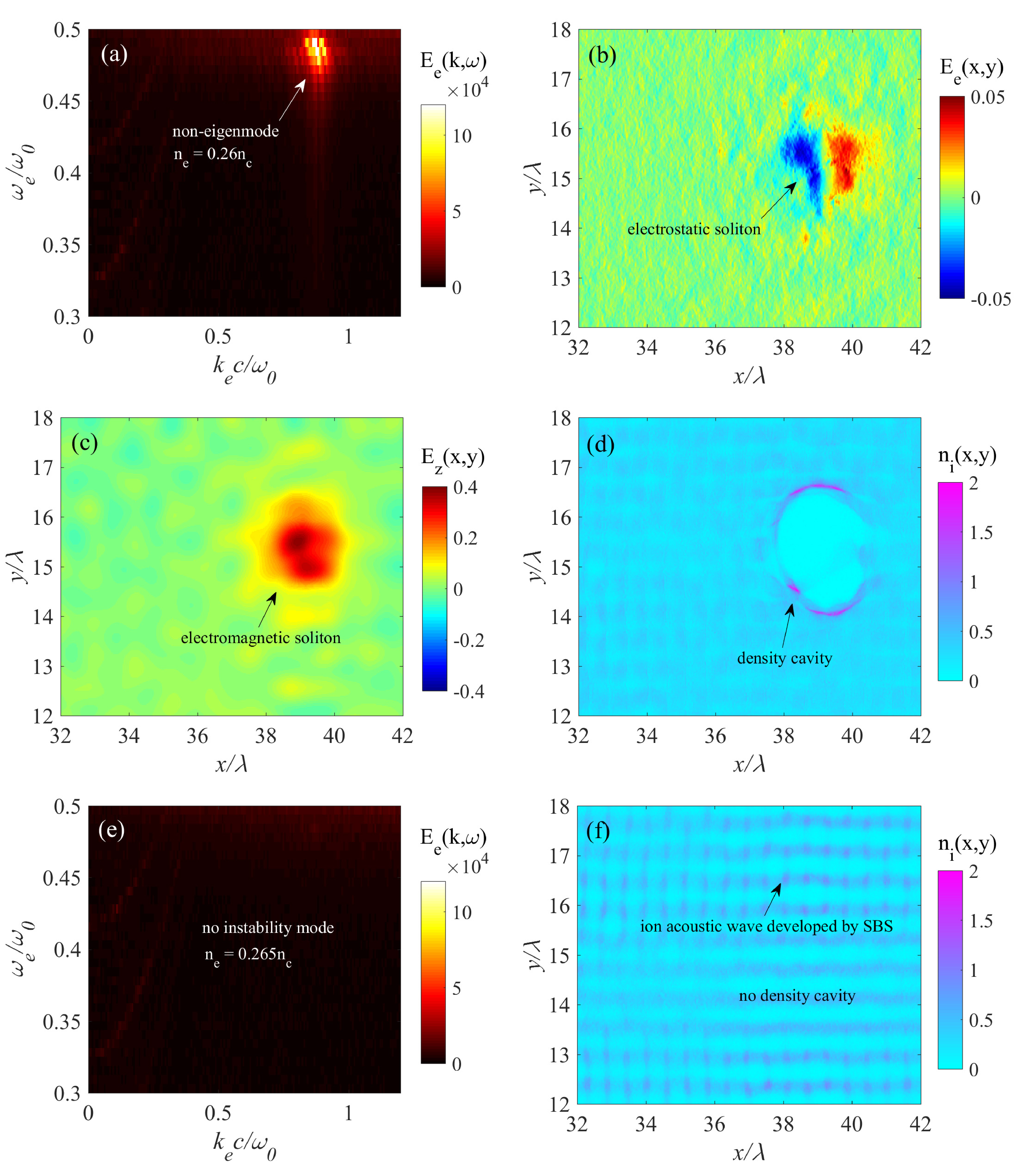}
        \end{overpic}
    \end{tabular}
\caption{ The plasma density is $n_e=0.26n_c$ for (a)-(d). (a) Distribution of the electrostatic wave in $(k_e, \omega_e)$ space obtained for the time window $[320, 480]\tau$ and transverse region $[14.4, 15.6]\lambda$. (b) Spatial distribution of electrostatic wave at $t=1850\tau$. (c) Spatial distribution of electromagnetic wave at $t=1850\tau$. (d) Spatial distribution of ion density at $t=1950\tau$. The plasma density is $n_e=0.265n_c$ for (e)-(f). (e) Distribution of the electrostatic wave in $(k_e, \omega_e)$ space obtained for the time window $[320, 480]\tau$ and transverse region $[14.4, 15.6]\lambda$. (f) Spatial distribution of the ion density at $t=1950\tau$. $E_e$ and $E_z$ are normalized by $m_e\omega_0c/e$, where $m_e$ and $e$ respectively are the electron mass and electron charge. $n_i$ is normalized by $n_c$.
    }
\end{figure*}

\begin{figure*}
\centering
    \begin{tabular}{lc}
        \begin{overpic}[width=0.78\textwidth]{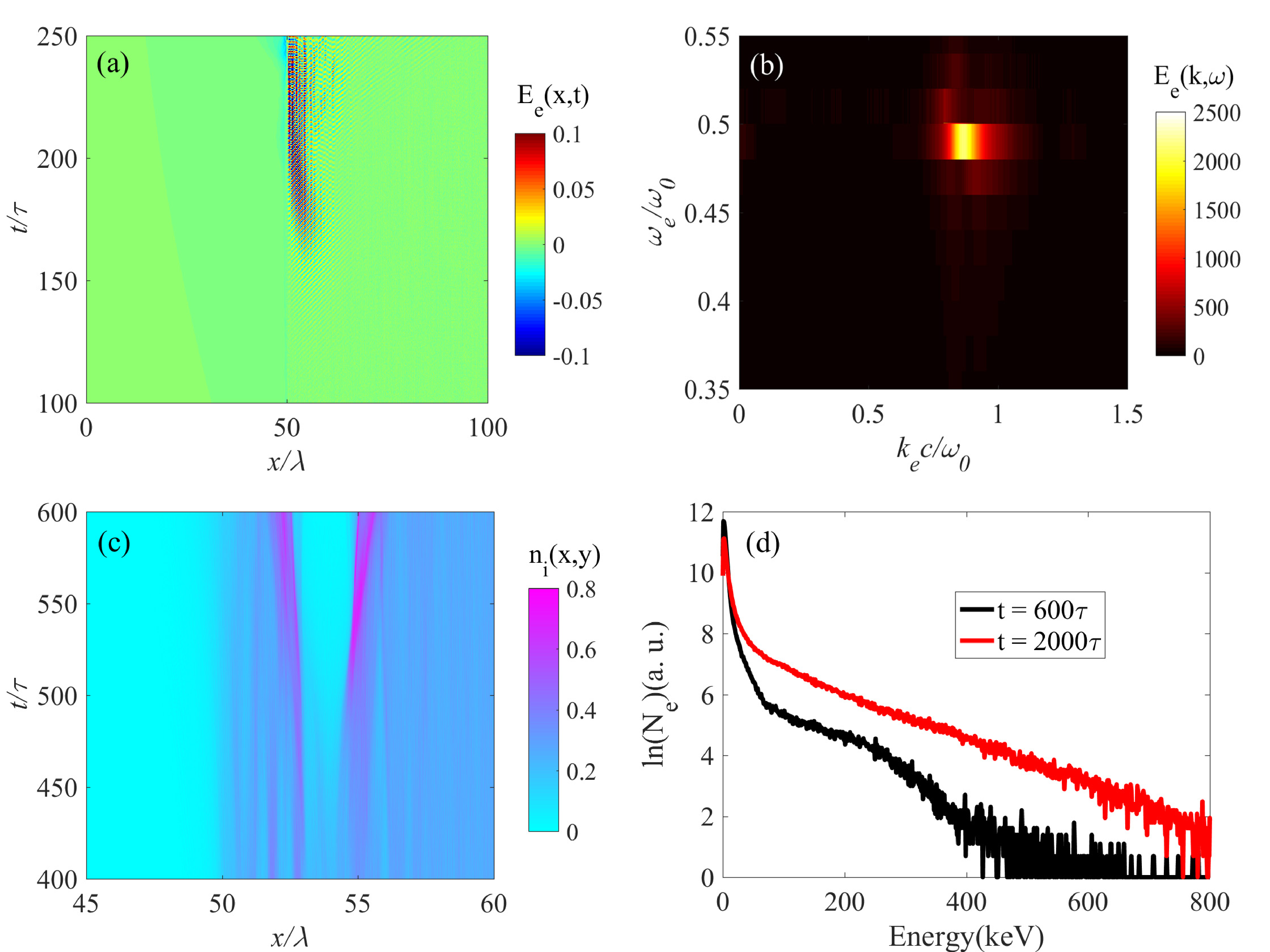}
        \end{overpic}
    \end{tabular}
\caption{ (a) The spatial-temporal distributions of electrostatic wave. (b) Distributions of the electrostatic wave in $(k_e, \omega_e)$ space obtained for the time window $[150, 200]\tau$. (c) The spatial-temporal distributions of ion density. (d) Energy distributions of electrons at different times. $E_e$ and $n_i$ respectively are normalized by $m_e\omega_0c/e$ and $n_c$.
    }
\end{figure*}

\subsection{2D simulations for degenerate SRS in homogeneous plasma}

To further validate the linear development and nonlinear evolution of degenerate SRS in high-dimensionality with mobile ions, we have performed several two-dimensional (2D) simulations. The plasma occupies a longitudinal region from 25$\lambda$ to 125$\lambda$ and a transverse region from 5$\lambda$ to 25$\lambda$ with homogeneous density $n_e=0.26n_c$. The initial electron temperature is $T_e=100$eV. Ions are movable with mass $m_i=3672m_e$ and an effective charge $Z=1$. A s-polarized (electric field of light is perpendicular to the simulation plane) semi-infinite pump laser with a peak amplitude $a_0=0.05$ at focal plane $x=75\lambda$ is incident from the left boundary of the simulation box.

According to Eq. (4), we know that the incident laser with peak amplitude $a_0=0.05$ is sufficient to develop degenerate SRS at plasma density $n_e=0.26n_c$. The simulation results for plasma density $n_e=0.26n_c$ are displayed in Figs. 4(a)-4(d). Fourier transform of the electrostatic wave is taken for the time window $[320, 480]\tau$. We summate the Fourier spectrum along the transverse direction between $y=14.4\lambda$ and $y=15.6\lambda$, and show the distribution in Fig. 4(a). One can find a degenerate electrostatic mode around $k_ec=0.86\omega_0$ and $\omega_e=0.492\omega_0$. Note that the growth rate of SBS is about half of the degenerate SRS, therefore, SBS have little effect on the development of degenerate SRS. As discussed in Sec. 2, the group velocities of the electrostatic wave and electromagnetic wave associated with the degenerate SRS are zero. As a result, they will be trapped in the plasma. This is confirmed in our numerical simulation as shown in Figs. 4(b) and 4(c), the electrostatic wave and the concomitant electromagnetic wave form localized structures. The trapped light and electrostatic wave may cause the laser energy deficit in ICF related experiments \cite{zhao2019absolute}. The trapped waves expel the ions to form density cavity at later time $t=1950\tau$ as seen from Fig. 4(d). These plasma cavities subsequently affect the evolution of the degenerate SRS and SBS \cite{Weber2005,wu195202,riconda2006two}. Note that this density cavity is formed due to the degenerate SRS, which is different from the solitons generated by relativistic intensity lasers \cite{EsirkepovLow,BulanovSolitonlike,SentokuBursts,NaumovaFormation,wudong}.

The laser with peak amplitude $a_0=0.05<a_{th-n}=0.067$ is insufficient to develop degenerate SRS at $n_e=0.265n_c$. The simulation results under plasma density $n_e=0.265n_c$ are displayed in Figs. 4(e) and 4(f). The comparison between Figs. 4(a) and 4(e) indicates that the pump laser with peak amplitude $a_0=0.05$ fails to drive degenerate SRS at $n_e=0.265n_c$ when the amplitude threshold is not reached. Only the ion acoustic wave developed by SBS with wavenumber $k_ic=2k_0c=1.72\omega_0$ can be found in Fig. 4(f). And no density cavities has been formed under this conditions. These results further indicate that degenerate SRS is a seed for the subsequent nonlinear physical phenomena.

\subsection{1D simulations for degenerate SRS in inhomogeneous plasma}

To study the degenerate SRS in hot inhomogeneous plasma, we have performed a simulation for the inhomogeneous plasma $n_e=0.26\exp[(x-50)/1000]n_c$ with density range [0.26, 0.287]$n_c$. The plasma locates in $x=[50, 150]\lambda$, and two 50$\lambda$ vacuums are left on either side of the plasma. The initial electron temperature is $T_e=2$keV. Ions are movable with mass $m_i=3672m_e$. The ion charge and temperature respectively are $Z=1$ and $T_i=1$keV. The driving laser is a linearly-polarized semi-infinite pump lasers with a uniform amplitude $a_0=0.07$.

The spatial-temporal evolution of electrostatic wave is exhibited in Fig. 5(a). We find that a strong electrostatic wave has been developed at the front of plasma $x\lesssim60\lambda$ at $t=180\tau$. The electrostatic wave envelop is found to be stationary due to its group velocity $v_g=0$. Note that the spatial gradient has little effect on the development of degenerate SRS, by reason that the phase matching of the three waves is always satisfied in inhomogeneous plasma. Therefore, degenerate SRS is an absolute instability. Figure 5(b) shows the distribution of electrostatic wave in $(k_e, \omega_e)$ space, where one can find a spectrum around $\omega_e=0.499\omega_0$. This result further validates that the frequency of degenerate electrostatic wave is independent of plasma density and electron temperature. The electrostatic and electromagnetic wave trapped in plasma will expel ions. From Fig. 5(c) we know that ion density cavity is gradually formed from $t=400\tau$ at the front of plasma. Large numbers of hot electrons are produced by the degenerate electrostatic field as shown in Fig. 5(d). As discussed above, the phase velocity of the degenerate electrostatic field is around 0.58$c$. The temperature of the electron hot tail at $t=2000\tau$ is around 141keV. The transmission rate of the pump laser though plasma is about 19.46\% at $t=2000\tau$, which indicate that degenerate SRS is an important pump energy loss mechanism in the laser plasma interactions as long as the laser intensity is higher than $10^{15}$W/cm$^2$.

\section{Summary}

In summary, we have shown theoretically and numerically that the degenerate SRS develops at plasma density $n_e>0.25n_c$ when the laser amplitude is larger than a certain threshold. The electrostatic wave produced by the degenerate SRS has a constant frequency $\omega_0/2$, which is no longer the non-degenerate eigen electron plasma wave $\omega_{pe}$. The phase velocity of the degenerate electrostatic wave is about 0.58$c$, which corresponds to the electron energy of 175keV. Therefore, super hot electrons can be produced via the development of the degenerate SRS. The trapped electromagnetic wave and electrostatic wave associated with this instability can drive density cavities in plasma. Our theoretical model is validated by PIC simulations. The degenerate SRS is an important pump energy loss mechanism in the laser plasma interactions as long as the laser intensity higher than $10^{15}$W/cm$^2$.

\section{Acknowledgement}

This work was supported by the Natural Science Foundation of Shanghai (No. 19YF1453200), the National Natural Science Foundation of China (Nos. 11775144 and 1172109), and the Strategic Priority Research Program of Chinese Academy of Sciences (Grant Nos. XDA25050800 and XDA25050100). The authors would like to acknowledge the OSIRIS Consortium, consisting of UCLA and IST (Lisbon, Portugal) for providing access to the OSIRIS 4.0 framework.

%\bibliographystyle{apsrev4-1}
%\bibliography{ref}

%merlin.mbs apsrev4-1.bst 2010-07-25 4.21a (PWD, AO, DPC) hacked
%Control: key (0)
%Control: author (72) initials jnrlst
%Control: editor formatted (1) identically to author
%Control: production of article title (-1) disabled
%Control: page (0) single
%Control: year (1) truncated
%Control: production of eprint (0) enabled
%

\end{document}